\documentstyle[prl,aps,multicol,epsfig,array]{revtex}

\begin{document}
\draft

\title{Spontaneous Branching of Anode-Directed Streamers 
between Planar Electrodes}
\author{Manuel Array\'as$^{1,2}$\cite{Madrid}, 
Ute Ebert$^1$ and Willem Hundsdorfer$^1$}
\address{
$^1$CWI, P.O.Box 94079, 1090 GB Amsterdam, The Netherlands, 
      }
\address{
$^2$Instituut--Lorentz, Universiteit Leiden, P.O.Box 9506, 2300 RA Leiden, 
The Netherlands }
\date{revised version from March 10, 2002 of paper submitted to PRL on Nov. 16, 2001}
\maketitle

\begin{abstract} 
  Non-ionized media subject to strong fields can become locally
  ionized by penetration of finger-shaped streamers.
  We study negative streamers between planar electrodes in a simple 
  deterministic continuum approximation. We observe that
  for sufficiently large fields, the streamer tip can split. 
  This happens close to the limit of ``ideal conductivity''.
  Qualitatively the tip splitting is due to a Laplacian instability
  quite like in viscous fingering. For future quantitative
  analytical progress, our stability analysis of planar fronts 
  identifies the screening length as a regularization mechanism. 
\end{abstract}


\begin{multicols}{2}

Streamers commonly appear in dielectric breakdown
when a sufficiently high voltage is suddenly applied 
to a medium with low or vanishing conductivity.
They consist of extending fingers of ionized matter and
are ubiquitous in nature and technology \cite{Rai,Eddie}.
The degree of ionization inside a streamer is low, hence
thermal or convection effects are negligible. However, 
streamers are nonlinear phenomena due 
to the space charges inside the ionized body that modify 
the externally applied electric field.
While in many applications, streamers by a strongly non-uniform 
background electric field are forced to propagate towards 
the cathode through complex mixtures of gases
\cite{Eddie,Kuli,DBM}, we here investigate the basic phenomenon
of the primary anode-directed streamer in a simple non-attaching
and non-ionized gas and in a uniform background field as 
in the pioneering experiments of Raether 
\cite{Raether}. In previous theoretical work, it is implicitly 
assumed that streamers in a uniform background 
field propagate in a stationary manner \cite{Firsov,Dya,Raizer2}.
This view seems to be supported by previous simulations \cite{DW,Vit}.

In this paper we present the first numerical evidence that anode directed 
(or negative) streamers do branch even in a uniform background field 
and without initial background ionization in the minimal fully deterministic 
``fluid  model'' \cite{Rai,Firsov,Dya,Raizer2,DW,Vit}, 
if the field is sufficiently strong. We argue that this happens 
when the streamer approaches what we suggest to call the Lozansky--Firsov
limit of ``ideal conductivity'' \cite{Firsov}. 
The streamer then can be understood as an interfacial pattern with
a Laplacian instability \cite{Ute}, qualitatively similar
to other Laplacian growth problems \cite{Ivantsov}.
For future quantitative analytical progress, we identify the
electric screening length as a relevant regularization mechanism.
Our finding casts doubts on the existence of a stationary mode of
streamer propagation with a fixed head radius.

\end{multicols}

\begin{figure}
\label{fig1}
\begin{center}
\epsfig{figure=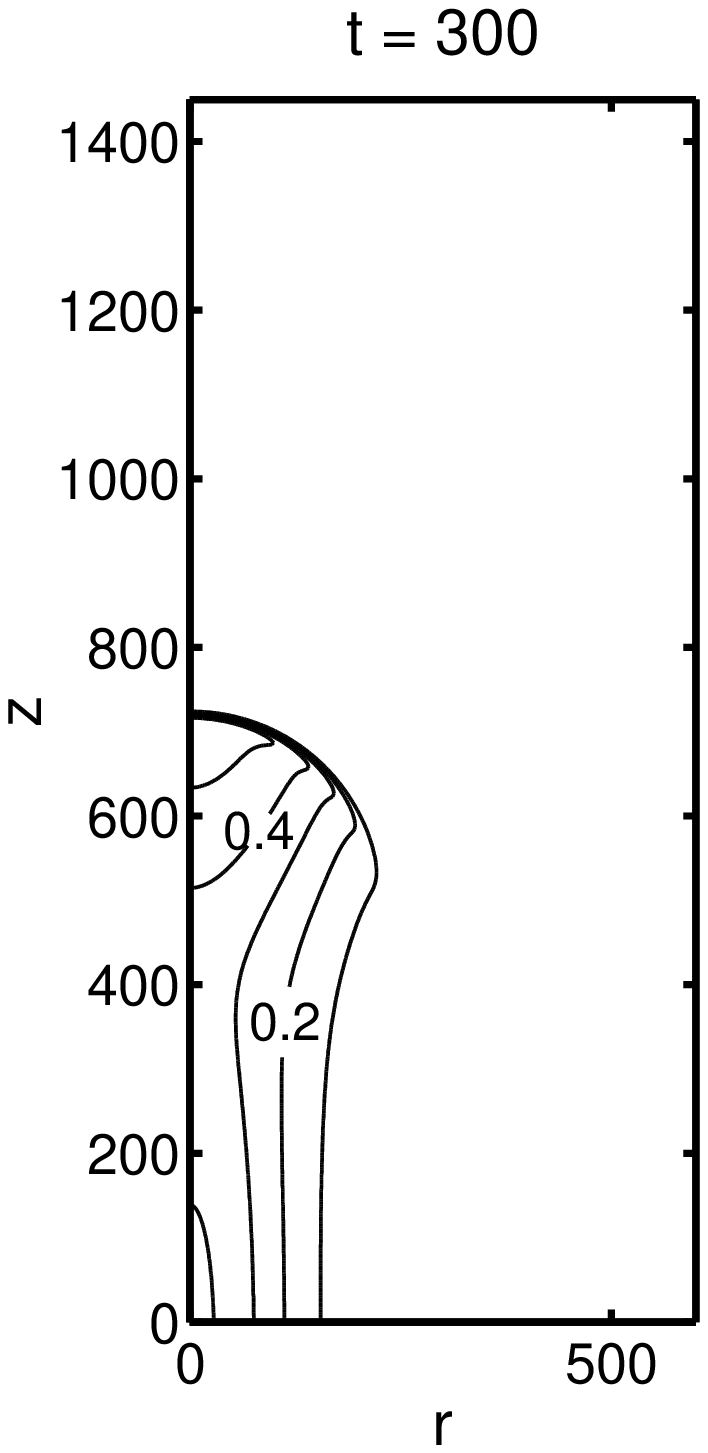,width=0.2\linewidth}
\epsfig{figure=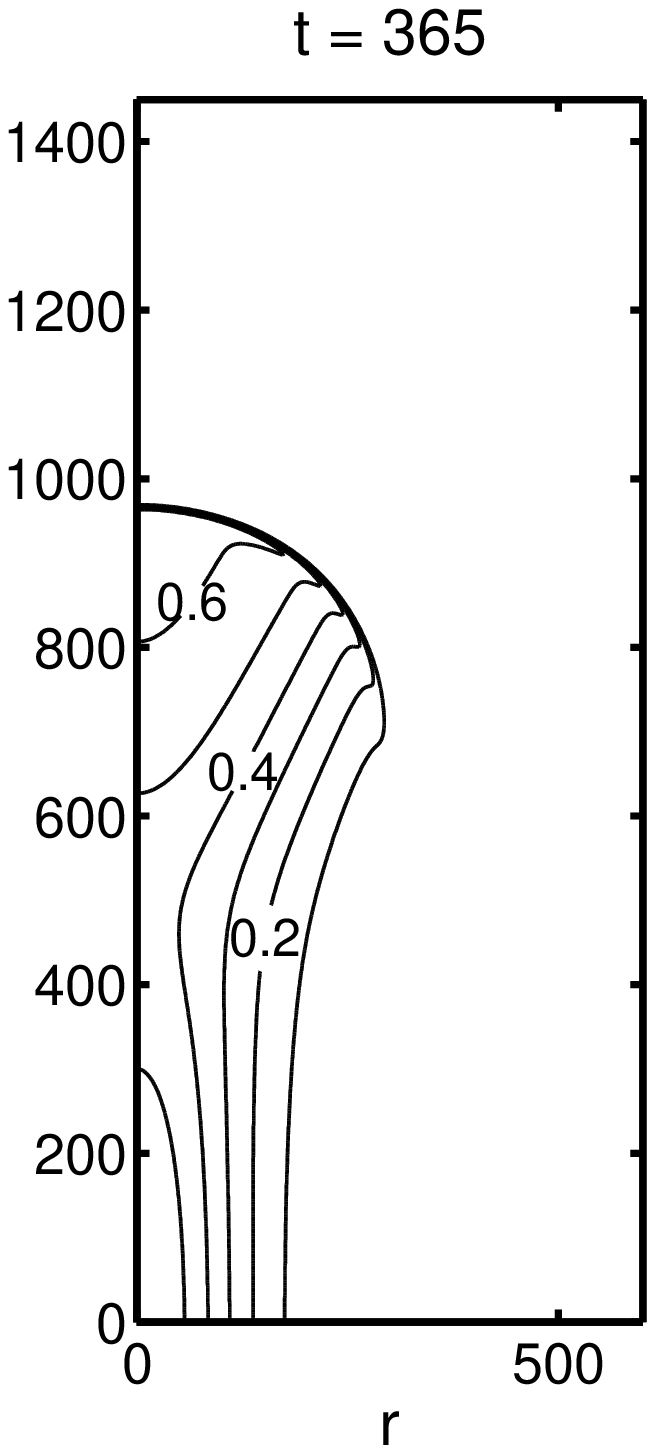,width=0.18\linewidth}
\epsfig{figure=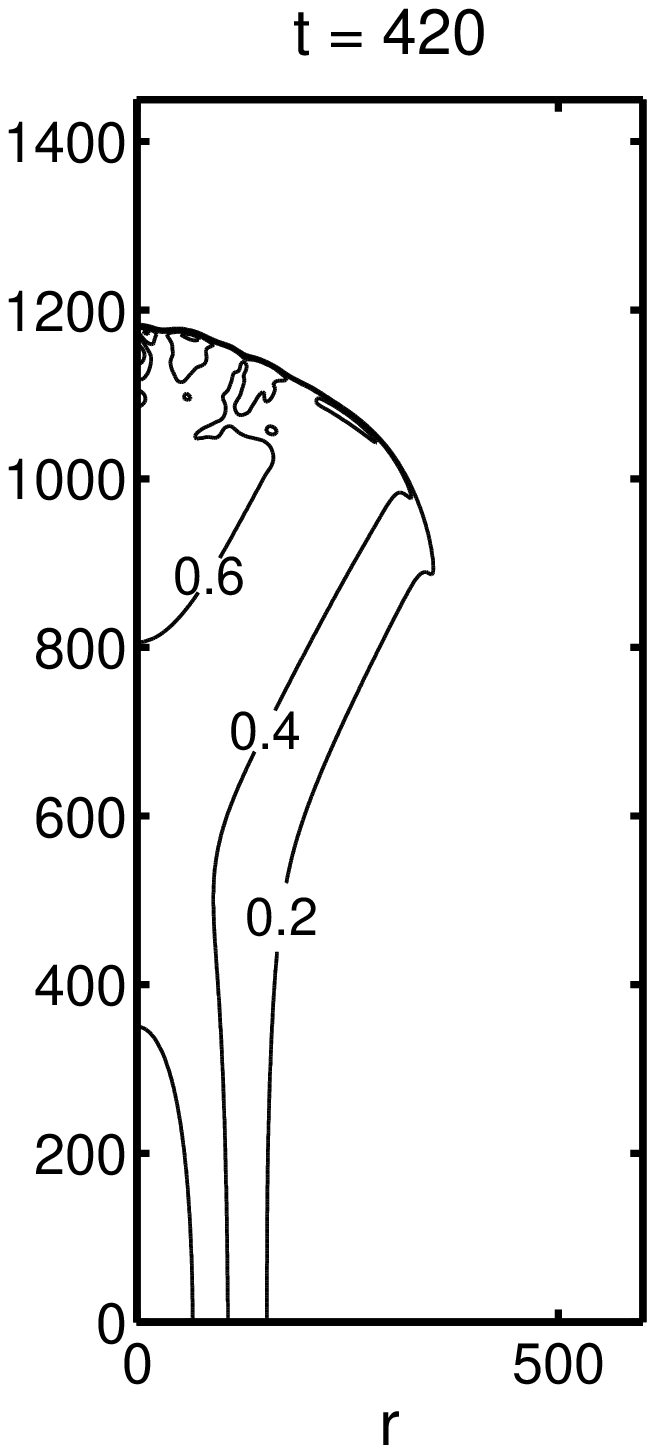,width=0.18\linewidth}
\epsfig{figure=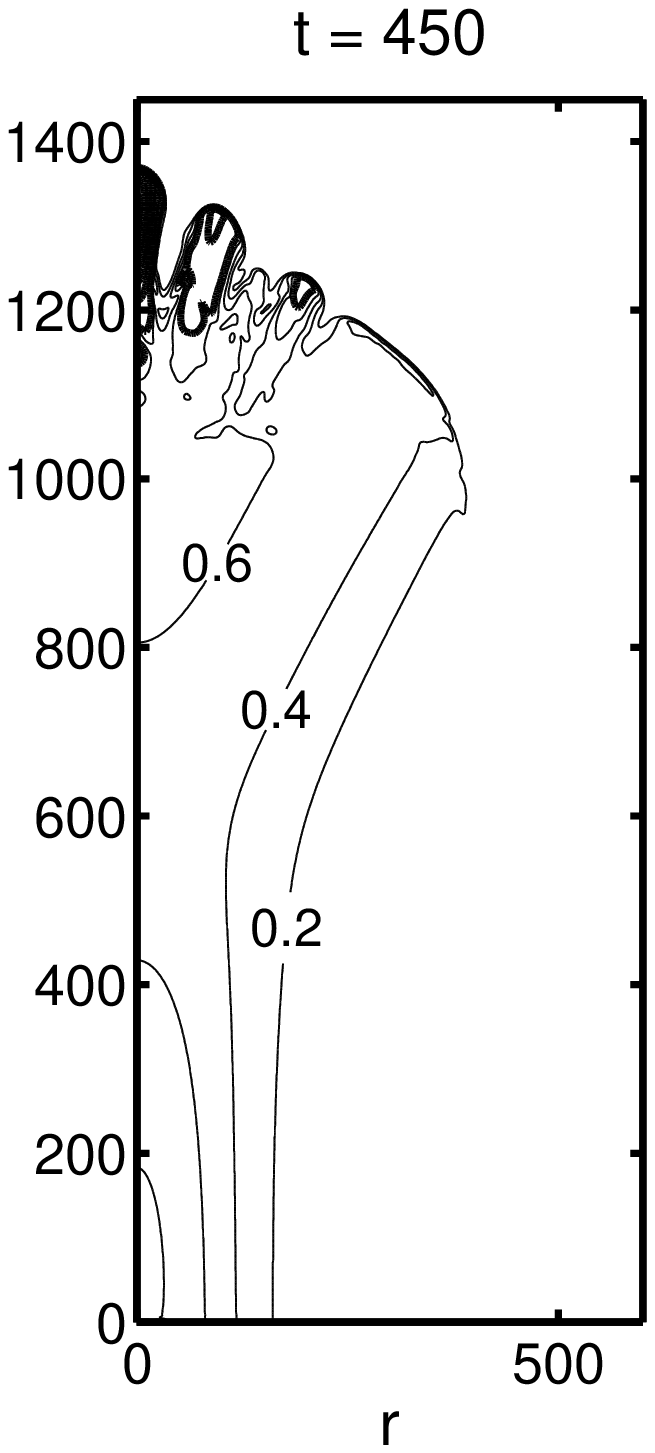,width=0.18\linewidth}
\caption[]{Evolution of spontaneous branching of anode directed streamers
in a strong homogeneous background field at times $t=300$, 365, 420 and 450.
Model, initial and boundary conditions are discussed in the text. 
The planar cathode is located at $z=0$ and the planar anode at $z=2000$ 
(shown is $0\le z\le 1400$). The radial coordinate extends from the origin 
up to $r=2000$ (shown is $0\le r\le 600$). The thin lines denote 
levels of equal electron density $\sigma$ with increments of 0.1 or 0.2 
as indicated by the labels. The thick lines denote the higher electron 
density levels 1., 2., 3., 4., 5.\ and 6. These high densities appear 
only at the last time step $t=450$ in the core of the new branches.}
\end{center}
\end{figure}

\begin{multicols}{2}

We investigate the {\it minimal streamer model}, i.e., a ``fluid
approximation'' with local field-dependent impact ionization reaction 
in a non-attaching gas like argon or nitrogen
\cite{Rai,Firsov,Dya,Raizer2,DW,Vit,Ute}.
In detail, the dynamics is as follows: \\
({\it i}) an impact ionization reaction in local field approximation: 
free electrons and positive ions are generated by impact of 
accelerated electrons on neutral molecules
$\partial_\tau n_e+\nabla_{\bf R}\cdot{\bf j}_e=
\partial_\tau n_i+\nabla_{\bf R}\cdot{\bf j}_i=
|\mu_e{\cal E}n_e|\;\alpha_0\;\alpha(|{\cal E}|/E_0)$;
$n_{e,i}$ and ${\bf j}_{e,i}$ are particle densities or currents
of electrons or ions, respectively, and ${\cal E}$ is the electric field;
in all numerical work, we use the Townsend approximation 
$\alpha_0\;\alpha(|{\cal E}|/E_0)=\alpha_0\;\exp(-E_0/|{\cal E}|)$ 
with parameters $\alpha_0$ and $E_0$ for the effective cross-section.
\\
({\it ii}) drift and diffusion of the charged particles in the local
electric field ${\bf j}_e=-\mu_e{\cal E}n_e-D_e\nabla_{\bf R}n_e$, 
where in anode-directed streamers the mobility of the ions actually 
can be neglected because it is more than two orders of magnitude 
smaller than the mobility $\mu_e$ of the electrons, so ${\bf j}_i=0$,
\\
({\it iii}) the modification of the externally applied electric field 
through the space charges of the particles according to the Poisson equation
$\nabla_{\bf R}\cdot{\cal E}=e(n_i-n_e)/\epsilon_0$.
It is this coupling between space charges and electric field 
which makes the problem nonlinear.

The natural units of the model are given by the ionization length 
$R_0=\alpha_0^{-1}$, 
the characteristic impact ionization field $E_0$, and
the electron mobility $\mu_e$ determining the velocity $v_0=\mu_e E_0$
and the time scale $\tau_0=R_0/v_0$. 
Hence we introduce the dimensionless coordinates \cite{Ute} 
${\bf r}={\bf R}/R_0$ and $t=\tau/\tau_0$, the dimensionless field
${\bf E}={\bf {\cal E}}/E_0$, the dimensionless electron
and ion particle densities $\sigma=n_e/n_0$ and $\rho=n_i/n_0$ 
with  $n_0=\varepsilon_0 E_0/(e R_0)$,
and the dimensionless diffusion constant $D=D_e/(R_0v_0)$. 
After this rescaling, the model has the form: 
\begin{eqnarray}
\label{1}
\partial_t\;\sigma \;-\; 
\nabla\cdot\left(\sigma\;{\bf E} + D\;\nabla\sigma\right)
&=& \sigma \; f(|{\bf E}|)~,
\\
\label{2}
\partial_t\;\rho \;
&=& \sigma \; f(|{\bf E}|)~,
\\
\label{3}
\rho - \sigma &=& \nabla\cdot{\bf E}~~,~~{\bf E}=-\nabla \Phi~,
\\
\label{5}
f(|{\bf E}|)=|{\bf E}|\;\alpha(|{\bf E}|)~
&\Big(=&|{\bf E}|\;e^{-1/|{\bf E}|}~\mbox{in sim.}\Big)~.
\end{eqnarray}

In the simulations presented here, a planar cathode is
located at $z=0$ and a planar anode at $z=2000$. The stationary
potential difference between the electrodes $\Delta\Phi=1000$ 
corresponds to a uniform background field ${\bf E} = -0.5 \;{\bf e}_z$
in the $z$ direction. For nitrogen under normal conditions with
effective parameters as in \cite{DW,Vit}, this corresponds to
an electrode separation of $\approx$ 5 mm and a potential
difference of $\approx$ 50 kV. The unit of time $\tau_0$
is $\approx3$ ps, and the unit of field $E_0$ is $\approx 200$ kV/cm.
We used $D=0.1$ which is appropriate for nitrogen, and we assumed
cylindrical symmetry of the streamer. The radial coordinate
extends from the origin up to $r=2000$ to avoid lateral
boundary effects on the field configuration.
As initial condition, we used an electrically neutral
Gaussian ionization seed on the cathode
\begin{equation}
\label{6}
\sigma(r,z,t=0)=\rho(r,z,t=0)
=10^{-6}\; e^{-(z^2+r^2)/100^2}. 
\end{equation} 
The parameters of our numerical experiment are essentially
the same as in the earlier simulations of Vitello {\it et al.}
\cite{Vit}, except that our background electric field is twice
as high; the earlier work had 25 kV applied over a gap of 5 mm.
This corresponded to a dimensionless background field of 0.25,
and branching was not observed.

In Fig.~1 we show the electron density levels at four time
steps of the evolution in the higher background field of 0.5.
We observe that at time $t=420$, the streamer develops 
instabilities at the tip. At time $t=450$, these instabilities 
have grown out into separate fingers. Because of the imposed 
cylindrical geometry, the further evolution after branching 
ceases to be physical. On the other hand, the main effect
of the unphysical symmetry constraint is to suppress all linear
instability modes that are not cylindrically symmetric. Hence in 
a fully 3D system, the instability will develop even earlier 
than here. 

Further simulations show: 
$(a)$ branching does not occur in a system of the same size 
in the lower background field of 0.25, in agreement with \cite{Vit}.
$(b)$ Branching is not due to the proximity of the anode, 
since in a system with twice the electrode separation 
(with the anode at $z=4000$) and with twice the potential difference 
($\Delta\Phi=2000$) --- so with the same background field ---, 
the streamer branches in about the same way after about the same time 
and travel distance.
$(c)$ The phenomenon is not specific to the particular initial condition (5).
$(d)$ Branching does somewhat depend on the numerical 
discretization. A wider numerical mesh leads to a higher effective 
noise level; and the branching then is triggered somewhat earlier. 
$(e)$ Occasionally, 
we observe a different tip splitting mode. In Fig.~1 at time
$t=450$, the finger on the axis develops the strongest with 
$\sigma$ exceeding 6, while in the fingers off the axis,
$\sigma$ stays below 3. In the other branching mode, 
the first finger off the axis outruns the finger on the axis.

Before we discuss the physical nature of the instability, we explain
our numerical approach: we used uniform space-time grids with 
a spatial mesh of $1000 \times 1000$. 
The spatial discretization is based on local mass balances.
The diffusive fluxes are approximated in standard fashion with 
second order accuracy. For the convective fluxes a third order upwind-biased 
formula was chosen to reduce the numerical oscillations that are common
with second order central fluxes. Such oscillations can be completely
avoided, e.g., by flux-limiting, but preliminary tests showed
that the upwind-biased formula already gives sufficient numerical
monotonicity and is much faster. Time stepping is based on an explicit
linear 2-step method, where at each time step the Poisson equation is
solved by the {\sc FISHPACK} routine. References for these procedures
can be found in \cite{Wess}.

\end{multicols}

\begin{figure}
\label{fig2}
\begin{center}
\epsfig{figure=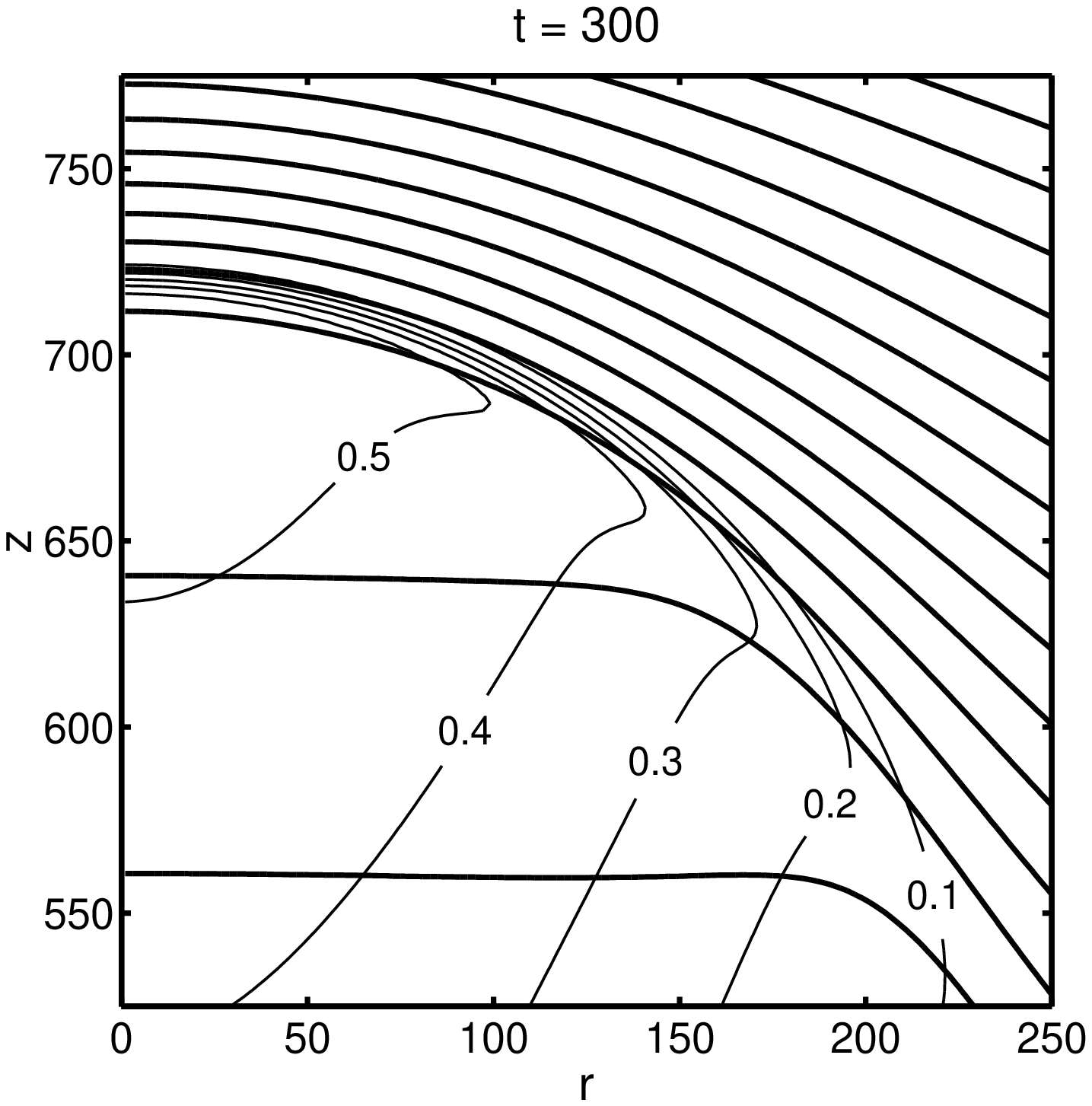,width=0.40\linewidth}
\epsfig{figure=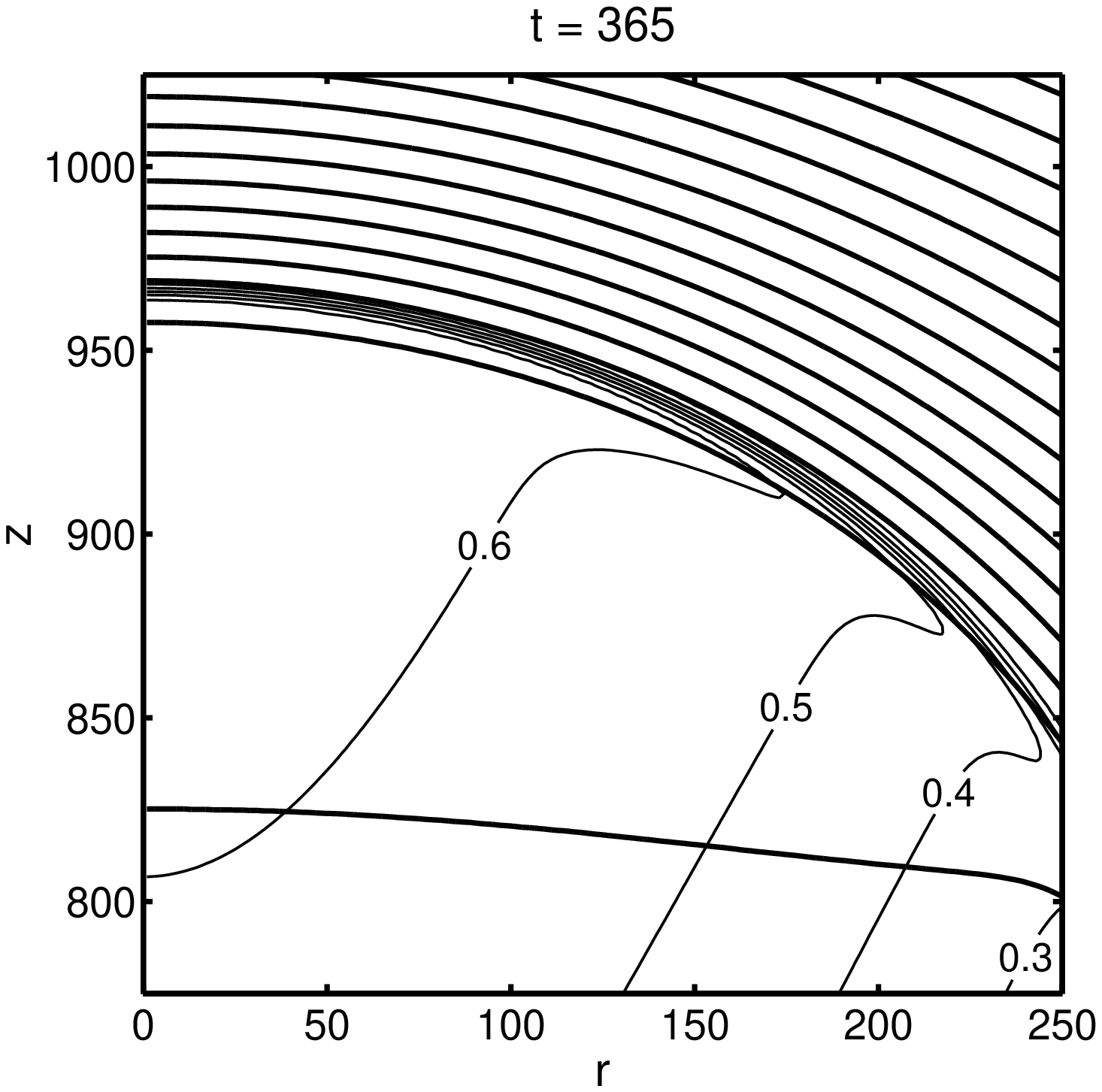,width=0.41\linewidth}
\caption[]{A zoom into the head of the streamer from Fig.\ 1 
at the first two time steps. The aspect ratio is equal and 
the axis scaling identical at both times. The thin lines
are the levels of equal electron density as in Fig.\ 1.
The thick lines are electrical equipotential lines in steps 
of $\Delta\Phi=12$.}
\end{center}
\end{figure}

\begin{multicols}{2}

To understand now why and at which stage the streamer develops 
a tip splitting instability, in Fig.\ 2, we zoom into the streamer 
head. Shown are the first two time steps from Fig.~1 with 
the electron density levels again as thin lines, and additionally 
with the equipotential lines as thick lines. 
One observes that during the temporal evolution prior to branching, 
both the curvature and the thickness of the ionization front decrease. 
So the width of the front becomes much smaller than its radius of 
curvature, and an interface approximation becomes increasingly justified. 
The electric field inside the streamer head also decreases, so that
the ionization front more and more coincides with an equipotential surface. 
In summary, the ionization front evolves towards a weakly curved and
almost equipotential moving ionization boundary. At the same time,
the electric field immediately ahead of the streamer increases.

We argue now that a transient stage of an approximately equipotential
and weakly curved ionization boundary leads to tip splitting. Conversely, 
we argue that tip splitting in the lower background field of 0.25 is 
not observed within the presently and previously \cite{Vit} investigated 
gap lengths because the transient stage of Fig.\ 2 is not reached 
before the streamer reaches the anode. 

In fact, the streamer in Fig.\ 2 approaches the limit of ``ideal 
conductivity'': the conducting body has $\Phi\approx {\rm const.}$, 
while in the non-ionized 
region $\nabla^2\Phi=0$ due to the absence of space charges.
The boundary between the two regions moves approximately with the 
drift velocity $v_f\propto\nabla\Phi$ or with the diffusion
corrected velocity $v_f\propto\nabla\Phi\;
\left(1+2\sqrt{D\;\alpha(|\nabla\Phi|)/|\nabla\Phi|}\right)$ 
\cite{Ute}. Our simulations are the first numerical 
evidence that the ``ideal conductivity'' limit can be approached 
within our model.

This limit of ideally conducting streamers in an electric field 
that becomes uniform far ahead of the front was studied by Lozansky
and Firsov 
\cite{Firsov}. They realized that uniformly propagating paraboloids 
of arbitrary radius of curvature are solutions of this problem. 
They did not realize that these paraboloids are mathematically equivalent 
to the Ivantsov paraboloids \cite{Ivantsov} of dendritic growth found 
earlier.
The uniformly propagating Ivantsov paraboloids in the early 80'ies were 
identified as dynamically unstable. This is generally the case for
such so-called Laplacian growth problems without a regularization mechanism.
Since ideally conducting streamers also pose such 
a Laplacian growth problem \cite{Ute}, the dynamical instability 
of the structure shown in Fig.\ 2 can be expected,
and it actually occurs as can be seen in Fig.\ 1. This
explains qualitatively why tip splitting occurs.

For a quantitative analysis, a system specific regularization
mechanism has to be found \cite{Ute,Ivantsov}. 
Its identification is intricate because negative streamer fronts are
so-called pulled fronts whose dynamics is dominated by the leading edge
rather than the nonlinear interior of the front \cite{pulled}. Therefore
standard methods like the pertubative derivation
of a moving boundary approximation for the model (\ref{1})--(\ref{5}) 
does not work \cite{MBA}. (Pulling also implies
that standard numerical methods with adaptive grids are inefficient.)
However, the ionization front has two intrinsic length scales, 
a diffusion length and an electric screening length. We therefore
explore the approximation of $D=0$. It is smooth for the velocity of
negative fronts \cite{Ute} and eliminates the leading edge, 
and hence suppresses the pulled nature of the front.
Rather the front becomes a shock front for the electron 
density, while the intrinsic length scale of the electric screening 
layer behind the shock remains.

As a first step to understand the short wave length regularization 
of perturbations due to this screening length,
we have investigated the transversal instability modes
of a planar ionization front in the limit $D=0$ in a field
that approaches the uniform limit ${\bf E}=-E_\infty\;{\bf e}_z$
far ahead of the front. The planar unperturbed front propagates 
with velocity $v=E_\infty$, which equals the drift velocity of the 
electrons precisely at the shock front. The implicit analytical 
front solution can be found in \cite{Ute}. 
In a comoving frame $\xi=z-vt$, we denote it by 
$\big(\;\sigma_0(\xi), \rho_0(\xi), \Phi_0(\xi)\;\big)$.
The Fourier components $\big(\;\tilde\sigma_k, \tilde\rho_k, 
\tilde\Phi_k\;\big)$ of a transversal linear perturbation are 
defined through 
\begin{equation}
\sigma=\sigma_0(\xi)+\int dk\;\tilde\sigma_k(\xi)\;e^{ikx+st}+\ldots
~~~\mbox{etc.}
\end{equation}
For the derivation of the boundary conditions on the shock front, 
it is more convenient to write a single Fourier component as
$\sigma=\sigma_0\left(\xi-e^{ikx+st}\right)+\sigma_k(\xi)\;e^{ikx+st}+\ldots$.
With this ansatz and the auxiliary field $\psi_k=\partial_\xi\phi_k$, 
the Fourier components solve the inhomogeneous equation
\begin{eqnarray}
\label{Matr1}
&&\partial_\xi \left(
\begin{array}{c}
\sigma_k \\ 
\rho_k \\ 
\psi_k \\ 
\phi_k
\end{array}\right)={\bf M}_{s,k}\cdot
\left(\begin{array}{c}\sigma_k\\ \rho_k\\ \psi_k\\ \phi_k\end{array}\right)
- \left(\begin{array}{c}s\;\partial_\xi\sigma_0/(v+E_0)\\
s\;\partial_\xi\rho_0/v\\ E_0k^2\\0 
        \end{array}\right) ,
\\
&&~~~ \nonumber \\
\label{Matr2}
&&{\bf M}_{s,k}=\left(\begin{array}{cccc}
  \frac{s+2\sigma_0-f(E_0)-\rho_0}{v+E_0}&
  \frac{-\sigma_0}{v+E_0}&
  \frac{\partial_\xi\sigma_0-\sigma_0 f'(E_0)}{v+E_0}
  &0 \\
  -f(E_0)/v & s/v & -\sigma_0 f'(E_0)/v & 0 \\
  1 & -1 & 0 & k^2 \\
  0 & 0 & 1 & 0
\end{array}\right) .
\nonumber
\end{eqnarray}
The boundary conditions at the shock $\xi=0$ can be obtained from the
analytical solution in the non-ionized area, and from the boundedness
of the charge densities:
\begin{equation}
\label{Init}
\left(\begin{array}{c}\sigma_k\\ \rho_k\\ \psi_k\\ \phi_k\end{array}\right)
\stackrel{\xi\uparrow0}{\longrightarrow}
\left(\begin{array}{c}f'(v)/(1+s/f(v))\\ 0\\ 1\\ 
(vk-s)/(sk)\end{array}\right) .
\end{equation}
The other boundary conditions are obtained by imposing that at
$\xi\to-\infty$ the electric field decays and the densities become constant. 

These equations together with the boundary conditions define an
eigenvalue problem for $s=s(k,v)$ with $v=E_\infty$. It can be solved
numerically by shooting from $\xi=0$ towards $-\infty$. In agreement
with analytical limits --- details will be given elsewhere ---, we find
\begin{eqnarray}
\label{sk}
s(k)&=&\left\{\begin{array}{ll}
|E_\infty| \;k &~~\mbox{ for }k\ll\alpha(|E_\infty|)/2\\ 
|E_\infty|\;\alpha(|E_\infty|)/2 &~~\mbox{ for }k\gg\alpha(|E_\infty|)/2 
\end{array}\right. .
\end{eqnarray}
This means that the electric screening length $1/\alpha(|E_\infty|)$
does regularize the instability of short wave length
perturbations from linear growth in $k$ to the saturation 
value $s(k)=|E_\infty|\;\alpha(E_\infty)/2$. A small 
positive growth rate remains, but the analytical derivation of 
(\ref{sk}) hints to the unconventional possibility that sufficiently 
curved fronts actually are stable to short wave length perturbations.
This question is presently under investigation. If true, it would
identify a most unstable wave length determining the width of the 
fingers that emerge after tip splitting.

In conclusion, we have presented numerical evidence that 
anode-directed streamers in a sufficiently strong, but uniform field 
can branch spontaneously even in a fully deterministic fluid model. 
We have argued that this happens when the streamer approaches 
the limit of ideal conductivity. We have established a qualitative
mathematical analogy with tip splitting of viscous fingers through 
the concept of Laplacian growth, and we have analytically demonstrated
that the electric screening length leads to an unconventional
regularization. This opens the way to future quantitative analytical 
progress. 

{\bf Acknowledgement:} 
M.A.\ was supported by the EU-network ``Patterns, Noise, and Chaos'' 
and U.E.\ partially by the Dutch Science Foundation NWO.

\end{multicols}

\end{document}